\documentclass[	DIV=calc,%
							paper=a4,%
							fontsize=10pt,%
							twocolumn]{scrartcl}	 					
\usepackage{hyperref}

\usepackage{lipsum}													

\usepackage[english]{babel}										
\usepackage[protrusion=true,expansion=true]{microtype}				
\usepackage{amsmath,amsfonts,amsthm}					
\usepackage[pdftex]{graphicx}									
\usepackage[svgnames]{xcolor}									
\usepackage[small,labelfont=bf,textfont=it]{caption}	
\usepackage{epstopdf}												
\usepackage{subfig}													
\usepackage{booktabs}												
\usepackage{fix-cm}													
\usepackage{float}

\usepackage{sectsty}													
\allsectionsfont{
	\fontsize{11}{14}\usefont{OT1}{bch}{b}{n}
	}

\sectionfont{
	\color{DarkRed}\fontsize{13}{15}\usefont{OT1}{bch}{b}{n}
	}

\usepackage{fancyhdr}												
\usepackage{lastpage}	

\usepackage{lettrine}
\newcommand{\initial}[1]{%
     \lettrine[lines=3,lhang=0.3, nindent=0em]{
     				\color{DarkRed}
     				{\textsf{#1}}}{} }

\usepackage{titling}															

\newcommand{\HorRule}{\color{Grey}
									  	\rule{\linewidth}{1pt}\vspace{.2in}%
										}

\pretitle{\vspace{0in} \begin{flushleft} \HorRule \fontsize{32}{40} \usefont{OT1}{phv}{b}{n} \color{DarkRed} \selectfont }
\title{Quantum Computing @ MIT: \\ {\Huge The Past, Present, and Future of the Second Revolution in Computing}}					
\posttitle{\par\end{flushleft}\vskip 0.5em}

\preauthor{\begin{flushleft}
					\large \lineskip 0.5em \usefont{OT1}{phv}{b}{sl} \color{Black}}
\author{Francisca Vasconcelos\thanks{francisc@mit.edu}, }											
\postauthor{\footnotesize \usefont{OT1}{phv}{m}{sl} \color{Black} 
					Massachusetts Institute of Technology, Departments of EECS and Physics					
					\par\end{flushleft}\vspace{.2in}\HorRule \vspace{-.3in}}

\date{}																				

\begin{document}

\maketitle

\thispagestyle{empty} 	

\initial{E}\textbf{very school day, hundreds of MIT students, faculty, and staff file into 10-250 for classes, seminars, and colloquia. However, probably only a handful know that directly across from the lecture hall, in 13-2119, four cryogenic dilution fridges, supported by an industrial frame, endlessly pump a mixture of helium-3 and helium-4 gases to maintain temperatures on the order of 10mK. This near-zero temperature is necessary to effectively operate non-linear, anharmonic, superconducting circuits, otherwise known as qubits. As of now, this is one of the most widely adopted commercial approaches for constructing quantum processors, being used by the likes of Google, IBM, and Intel. At MIT, researchers are working not just on superconducting qubits, but on a variety of aspects of quantum computing, both theoretical and experimental. \vspace{0.05in}\\ 
\indent In this article we hope to provide an overview of the history, theoretical basis, and different implementations of quantum computers. In Fall 2018, we had the opportunity to interview four MIT faculty at the forefront of this field -- Isaac Chuang, Dirk Englund, Aram Harrow, and William Oliver -- who gave personal perspectives on the development of the field, as well as insight to its near-term trajectory. There has been a lot of recent media hype surrounding quantum computation, so in this article we present an academic view of the matter, specifically highlighting progress being made at MIT.
}

\begin{figure}[ht!]
    \centering
    \includegraphics[width=.92\columnwidth]{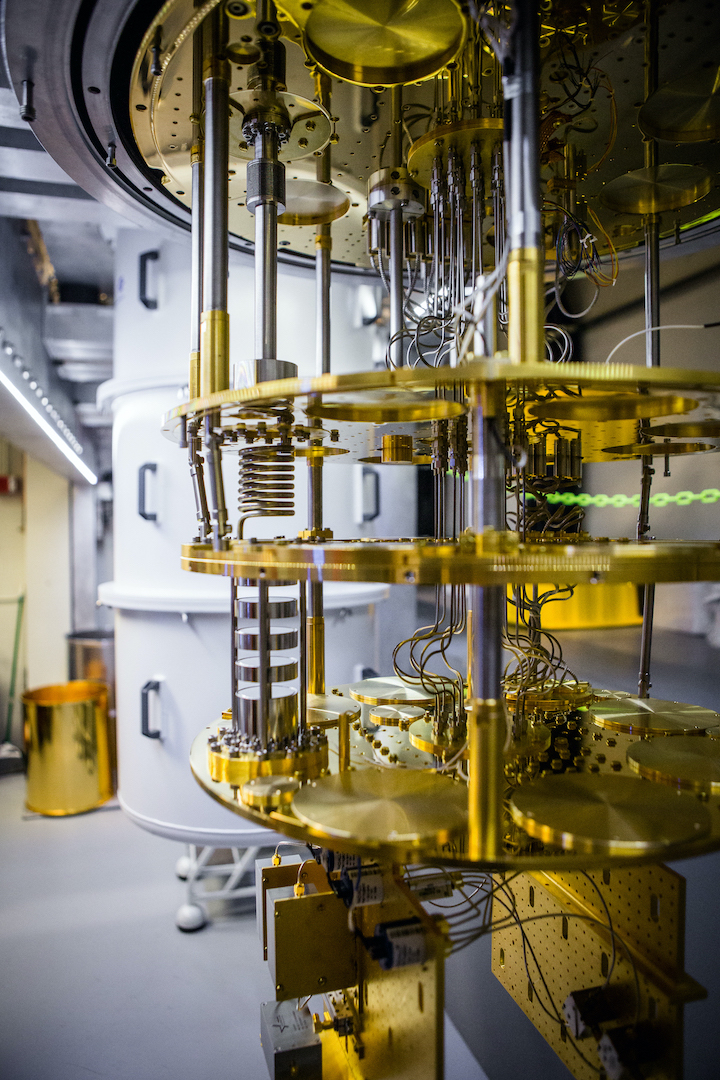}
    \caption*{Bluefors dilution refrigerators used in the MIT Engineering Quantum Systems group to operate superconducting qubits at near-zero temperatures. Credit: Nathan Fiske}
\end{figure}

\newpage
\section*{Quantum Computers: A Brief History}
The notion of a quantum computer was first introduced by Caltech Professor (and MIT alumnus) Richard Feynman in his 1960 \textit{``There's Plenty of Room at the Bottom"} lecture~\cite{feynman}, suggesting the use of quantum effects for computation. However, it was not until the late 1970s that researchers truly began exploring the idea. By May 1980, Paul Benioff -- then a researcher at the Centre de Physique Théorique, CNRS -- published the first article on quantum computation, \textit{``The computer as a physical system: A microscopic quantum mechanical Hamiltonian model of computers as represented by Turing machines"}, in the Journal of Statistical Physics~\cite{benioff1980computer}. In the same year, Russian mathematician and Steklov Mathematical Institute faculty, Yuri Manin published the book \textit{``Computable and Uncomputable"}~\cite{manin1980computable}, motivating the development of quantum computers. In 1981, MIT held the very first conference on the Physics of Computation at the Endicott House~\cite{fredkin1982papers}, where Benioff described computers which could operate under the laws of quantum mechanics and Feynman proposed one of the first models for a quantum computer. Clearly, MIT had roots in the field from its conception.

In 1992, David Deutsch and Richard Josza, proposed the Deutsch-Josza algorithm~\cite{deutsch1992rapid}. Although the algorithm in itself was not very useful, a toy problem some may say, it was one of the first demonstrations of an algorithm that could be solved far more efficiently on  a quantum computer than on a classical (or non-quantum) computer. This idea, that quantum computers may exhibit a better computational complexity than classical computers, is known as \textit{\color{DarkRed}quantum advantage}. In the years that followed, more work was done in the domain of quantum algorithms and in 1994 one of the biggest results in the field was achieved. Current MIT Professor of Applied Mathematics, Peter Shor, who was then working at Bell Labs, proposed the now well-known \textit{\color{DarkRed}Shor's algorithm}~\cite{shor1994algorithms}. This quantum algorithm reduced the runtime of integer number factorization from exponential to polynomial time. While this may seem like a fairly abstract problem, the assumption that integer factorization is hard is the foundation of RSA security, the primary cryptosystem in use today. As the first algorithm to demonstrate significant quantum advantage for a very important problem without obvious connection to quantum mechanics (like quantum simulation), Shor's algorithm caught the attention of the government, corporations, and academic scientists. This created major interest in the field of quantum information and gave experimental researchers a strong justification to develop quantum computers from the hardware end. 

The years following the announcement of Shor's algorithm witnessed significant developments in quantum computation, from both the theoretical and experimental perspectives. In 1996, Lov Grover of Bell Labs developed the quantum \textit{\color{DarkRed}Grover's algorithm}~\cite{grover}, which provides a quadratic speedup for search in unstructured problems, using a technique now known as amplitude amplification. That same year, the US Government issued its first public call for research proposals in the domain of quantum information, signifying the growing interest in quantum computation. Additionally, David P. DiVincenzo, then at IBM Research, listed the five main requirements to realize a physical quantum computer~\cite{divincenzo1997topics}.  Among these included the challenges of isolating quantum systems from their noisy environments and accurately controlling unitary transformations of the system. 1997 witnessed the first publications of papers realizing physical gates for quantum computation, based on nuclear magnetic resonance (NMR)~\cite{cory1997ensemble, gershenfeld1997bulk}.  Three out of the five authors on these transformative papers (Isaac Chuang, Neil Gershenfeld, and David Cory) are current or previous MIT faculty. Additionally, that year, two new physical approaches to quantum computing were proposed, making use of majorana anyons~\cite{kitaev2003fault}  and quantum dots~\cite{loss1998quantum}. In 1998, the University of Oxford, followed shortly after by a collaboration between IBM, UC Berkeley, and the MIT Media Lab, ran the Deutsch-Jozsa algorithm on 2-qubit NMR devices~\cite{chuang1998experimental}.  These served as the very first demonstrations of an algorithm implemented on a physical quantum computer. This was soon followed by the development of a 3-qubit NMR quantum computer, a physical implementation of Grover's algorithm, and advances in quantum annealing. Thus, by the end of the twentieth century, researchers demonstrated the physical potential of quantum computing devices. 

The turn of the century would mark a transition of focus towards improving and scaling these devices. The early 2000s alone witnessed the scaling to 7-qubit NMR devices, an implementation of Shor's algorithm which could factor the number 15, the emergence of linear optical quantum computation, an implementation of the Deutsch-Jozsa algorithm on an ion-trap computer, and the first demonstration of the quantum XOR (referred to as the CNOT gate). The following years consisted of similarly remarkable developments in and scaling of quantum computation technology. In fact, this growth has been so impressive, that it is reminiscent of an exponential Moore's-Law-type growth in qubit performance.  This progress has resulted in excitement for quantum computation far beyond the realm of academia. As mentioned earlier, several large tech companies now have well-established research divisions for quantum technologies and the number of quantum startups seem to double each year. Furthermore, there has been a large increase in interest from both the government, with the approval of a \$1.2 billion Quantum Initiative Act in 2018 – one of the only bipartisan legislative acts passed in recent memory -- and the general public, with Google's recent announcement of quantum supremacy~\cite{arute2019quantum}. However, with this increased interest and the desire for easily-approachable explanations to complex research comes the tendency to hype the current state of the field. Thus, we interviewed four faculty at the forefront of academic research in quantum information and computation for their outlooks, to see if we could make sense of where all these quantum technologies are headed in the near- and long-term.

\newpage

\section*{Professor William Oliver -- Superconducting Quantum Processors}
William Oliver is the current Associate Director of the MIT Research Laboratory of Electronics (RLE), an Associate Professor of EECS, a Physics Professor of the Practice, and a Fellow of the MIT Lincoln Laboratory. He is also a Principal Investigator in the Engineering Quantum Systems Group (MIT campus) and the Quantum Information and Integrated Nanosystems Group (MIT Lincoln Laboratory), where he leads research on the materials growth, fabrication, design, and measurement of superconducting qubits.

\subsection*{How would Prof. Oliver explain his research to a non-expert?}
When we asked Professor Oliver to explain his research in general terms, he described superconducting qubits as ``artificial atoms." Like natural atoms, they have discrete quantum energy levels, in which transitions can be driven. One key difference, however, is that these \textit{\color{DarkRed}artificial atoms} are macroscopic electrical circuits comprising an Avogrado's number of actual atoms. Given that a qubit is a circuit, its energy levels can be engineered to be more optimally suited for quantum computation. 

At a high level, a \textit{\color{DarkRed}superconducting qubit} is an LC-circuit -- an inductor and capacitor in parallel -- like you might see in an E\&M course, such as 8.02. This type of circuit is a simple harmonic oscillator with a harmonic potential, meaning an equal energy spacing between all the energy levels. However, quantum bits are generally built from the ground and first excited energy levels, corresponding to the two possible bit states (0 and 1). In order to isolate these two lower energy levels from higher energy levels, quantum engineers make use of Josephson junctions as their inductors. At superconducting temperatures, \textit{\color{DarkRed}Josephson junctions} act as non-linear inductors, introducing an anharmonicity to the potential and creating unequal spacings between different energy levels. This allows the user to isolate the transition between the ground and first excited energy level from transitions to higher energy levels, solely by changing the microwave frequency at which the circuit is driven.

\begin{figure}[ht!]
    \centering
    \includegraphics[width=.92\columnwidth]{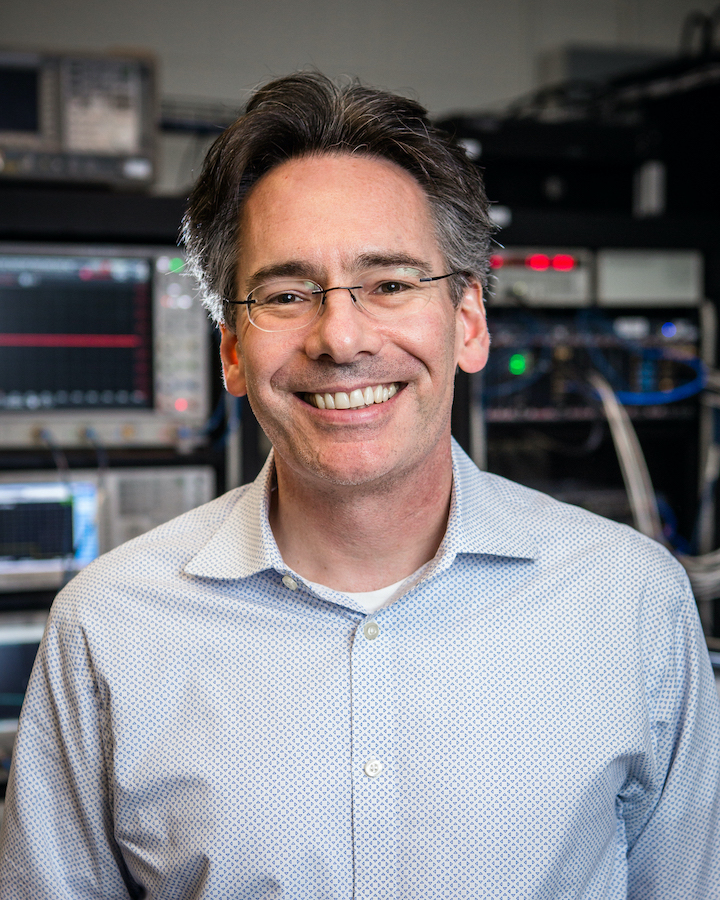}
    \caption*{Professor William Oliver, PI of the MIT Engineering Quantum Systems group. Credit: Nathan Fiske}
    \vspace{-.15in}
\end{figure}

\subsection*{How did Prof. Oliver get into superconducting qubit research?}

Professor Oliver did his PhD at Stanford in the group of Professor Yoshihisa Yamamoto. At the time, he was not working with superconducting qubits, but instead focused on electrons in two-dimensional electron gasses in semiconductor heterostructures. Specifically, he worked on recreating fundamental quantum optics experiments from the 1950s-80s, using electrons (fermions) instead of photons (bosons). Through this research, he learned a lot about different quantum properties, such as coherence (e.g., the phase coherence time of a quantum object), that are very important in modern quantum computing systems. He also became excited by the prospective use of solid-state systems to demonstrate quantum mechanical behavior. Although they were not long-lived enough for quantum computation, electrons in 2D electron gasses were sufficiently coherent to do the experiments he was interested in at the time. Professor Oliver claims that, during his PhD, he was motivated solely by “blue-sky physics” rather than engineering applications like quantum computation.

During his masters at MIT, Oliver met Professor Terry Orlando, who was developing the first superconducting persistent-current flux qubits. And, around that time, superconducting qubits were first demonstrated to behave like artificial atoms, but their coherence times were very short – less than 1ns. After his PhD, Oliver wanted to pivot from electron gases. He claims that there were a lot of interesting challenges in the field of superconducting qubits. Nobody knew if the technology would pan out. However, if it did eventually work, it would have a lot of nice features. In particular, superconducting qubits operate in the microwave regime and are fabricated in much the same way as transistors, making them lithographically scalable. Although their coherence times were short back then, the operation time was also very short. Professor Oliver felt that if he could extend the coherence time, then the number of operations performable within that coherence time would increase drastically. So, he decide to switch from semiconducting to superconducting quantum technologies. 

\begin{figure*}[ht!]
    \centering
    \includegraphics[width=\textwidth]{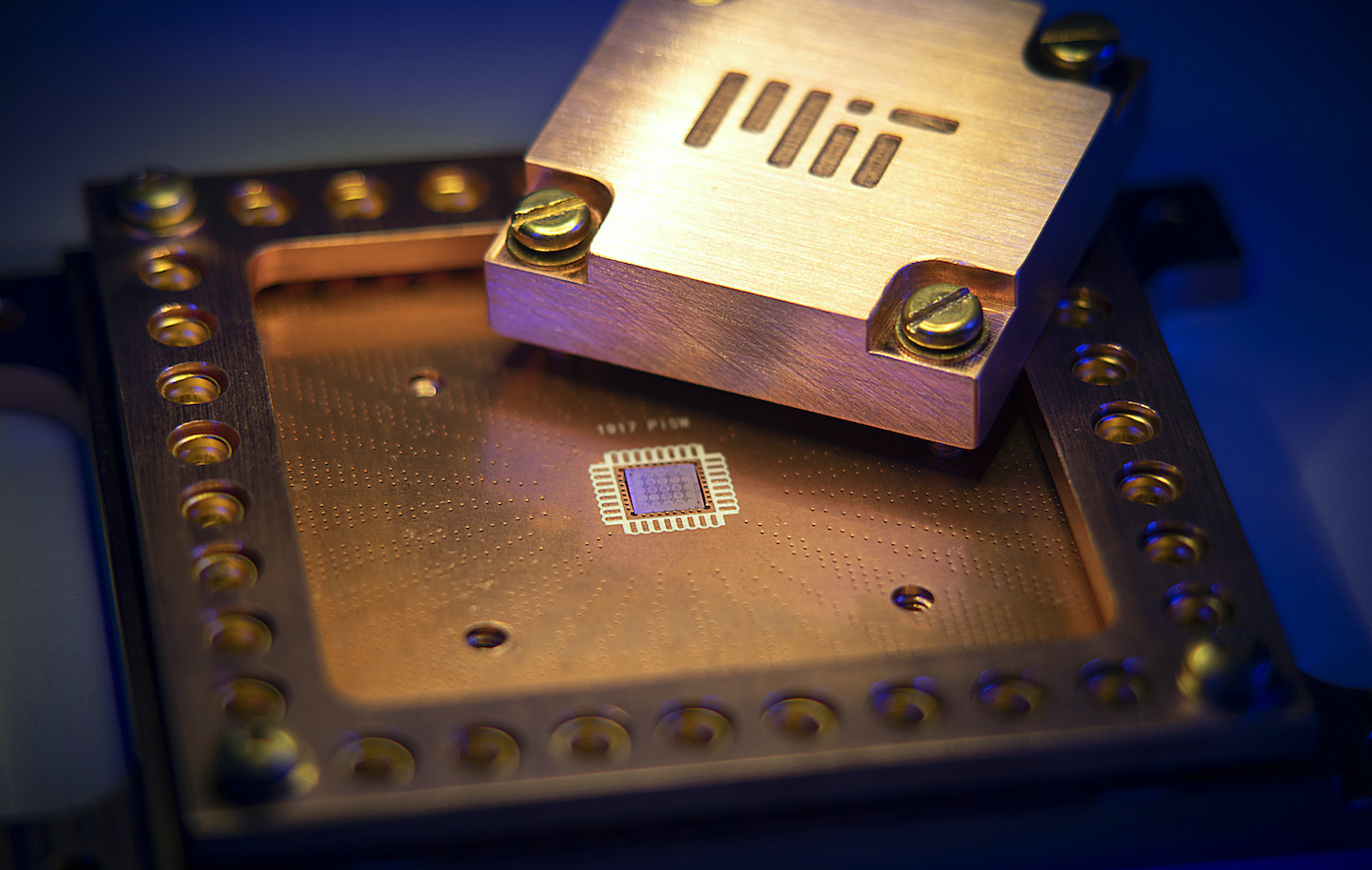}
    \caption*{A 16-qubit superconducting quantum processor and its
packaging, designed by researchers of the MIT Engineering Quantum Systems group
and fabricated at Lincoln Labs. Credit: MIT News}

\vspace{-.15in}
\end{figure*}

\subsection*{What were previous expectations in the field of superconducting qubits and have we met them?}

When Professor Oliver joined the field in 2003, coherence times were around 20ns, which was considerably shorter than competing trapped ion or nuclear-spin qubits. There were some quick results which came out, in which different groups achieved coherence times on the order of 500ns to even 4$\mu$s for single-qubit systems. However, these were generally one-off results. The best superconducting qubit during the mid-2000s was the persistent-current flux qubit, which was co-invented by Professor Orlando. The main challenge with this technology, however, was that the results were not reproducible. In a batch of 10 qubits, 2 might have great coherence times, 4 would not work, and 4 would be alright. At the same time, research was being done on other, more consistent types of superconducting qubits, such as the phase qubit developed by then NIST Fellow (and now UCSB Professor and current head of Google AI Quantum) John Martinis. However, this qubit faced the opposite problem: it would always work, but could never achieve a coherence time greater than 500ns. So, Professor Oliver, like many others in the field, dedicated his research to not only making superconducting qubits coherence times better, but also making them reproducible. 

Roughly 15 years later, Professor Oliver says that, in hindsight, nobody really expected that the field would undergo a Moore's Law-type scaling in improvement. We now, however, believe this is in fact the case, mostly due to significant contributions in design, materials, and fabrication. In fact, Oliver says that ``getting all three of these were very important. There was no one thing that we did that suddenly just made everything work. It was just a continual improvement, either designing qubits to be less sensitive to their noisy environment or by removing the sources of those noise (using material science or fabrication engineering)." He believes that the field has far exceeded its expectations from 2003-2005. People were hopeful that superconducting technology would become a leading platform for quantum computation, but it was largely disregarded as solely a laboratory curiosity in favor of other more advanced platforms of the time, like ion traps and NMR. Although expectations were low when the technology was in its youth, they are very high today.

\subsection*{Where are superconducting qubits headed?}

Superconducting qubits are currently in the process of transitioning from a scientific/laboratory curiosity to a technical reality. Professor Oliver believes that researchers need to bring a lot more to the field, in order to actually realize this technology. It is not sufficient to just demonstrate a one-qubit or two-qubit gate, but instead it necessary to build a reproducible system. In this effort, a new discipline is emerging, called \textit{\color{DarkRed}quantum engineering}, which bridges the gap between quantum science and conventional engineering. It covers a wide range of fields, including physics, mathematics, computer science, computer architecture, analog and digital design, control theory, digital signal processing, materials, fabrication, and more. In the long term, Professor Oliver believes all of these distinct domains need to come together to make superconducting qubit technology a viable technology. 

According to Oliver, one of the holy grails of modern quantum engineering research is to demonstrate a \textit{\color{DarkRed}logical qubit}, or redundant set of qubits that achieve higher performance in aggregate than individually. To achieve this, each individual physical qubit needs to have a sufficiently high operational \textit{\color{DarkRed}fidelity}, or ability to accurately perform gate operations. Currently, many superconducting groups can achieve single-qubit fidelities on the order of 99.95\%. However, coupled qubits have fidelities in the range of 95-99\%, with only a few groups (including Oliver's) able to achieve more than 99\%.  Professor Oliver believes that in order to demonstrate a logical qubit and run error-correcting codes with reasonable overhead, those fidelities should get better (consistently above 99.9\%, and the higher the better). He hopes that the \textit{\color{DarkRed}surface code}, a quantum error-correction scheme, will be implementable and demonstrated on superconducting devices of 17-49 qubits within the next 5 years. Furthermore, Oliver hopes that there will be a demonstration, in the 2D chip format, of aggregated qubits which are viable and last longer they otherwise would individually.  Demonstration of \textit{\color{DarkRed}quantum error correction} is a key milestone in the development of universal quantum computers, because it enables resilience through redundancy, enabling larger systems. 

The field is currently in the self-proclaimed \textit{\color{DarkRed}NISQ}, or Noisy Intermediate Scalable Quantum, Era in which there is access to small, noisy quantum processors. During the next 5-10 years, in parallel with the push towards making logical qubits and demonstrating quantum error-correction, Oliver believes that the field will also need to develop algorithms that make use of currently available NISQ devices. It is crucial to develop a quantum algorithm that gives a quantum advantage and addresses a useful, meaningful problem. 

He also hopes that all these algorithms and implementations will be commercializable. To date, quantum computing research has been largely funded by the government. Recently, a few companies, such as Google, IBM, Microsoft, and Intel, as well as a number of startups that have jumped in. While Oliver thinks this is great, he feels in order for the technology to prosper, these companies need to be able to sell something, generate revenue, and feed that into their next development cycle. In short, the ``virtuous cycle of development needs to kick in, which can only happen if we have a commercialized product." He claims that this is what drove the semiconductor industry, leading to the development of the personal computer. However, transistors and vacuum tubes could be used for amplifiers, radio receivers, and transmitters. They had applications completely unrelated to computing. Professor Oliver does not believe this is the case so far with superconducting qubits, with the only potential candidate being quantum sensing. Technology ``off-ramps" is something that he believes needs to be identified. In the meantime, the field also needs a NISQ algorithm that will take make these unprotected qubits useful, maybe as a co-processor with a classical computer. 30 years from now, Oliver hopes ``that we will have fault-tolerant quantum computers and use them to do amazing things that we never would have dreamt possible with classical computers. Until then, we also want to have NISQ-era algorithms that can also solve real-world problems."

\begin{figure}[t!]
    \centering
    \includegraphics[width=\columnwidth]{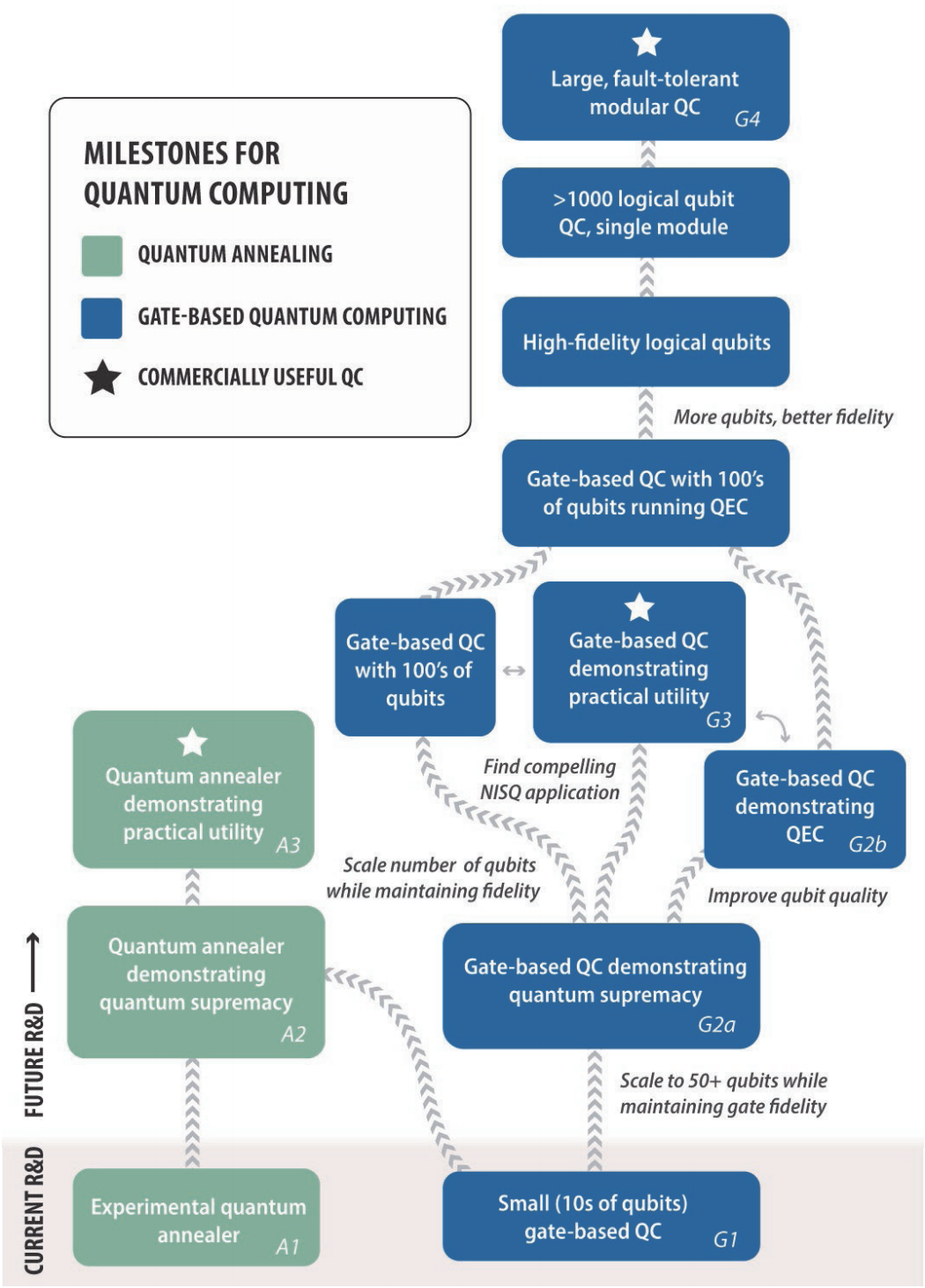}
    \caption*{An illustration of potential milestones of progress in quantum computing determined by the Committee on Technical Assessment of the Feasibility and Implications of Quantum Computing.~\cite{national2019quantum}}
    \vspace{-.15in}
\end{figure}

\subsection*{What kinds of problems will we use quantum computers to solve?}

Although the main national security driver for quantum computers is Shor's algorithm and cryptanalysis, Professor Oliver does not believe this will be the main commercial driver. In fact, he finds the commercial drivers (quantum chemistry, quantum materials, and optimization), much more exciting. ``These are just the thorny, sticky hard problems that we can't directly solve with classical computers because the number of degrees of freedom is too large and the number of requirements are too large. We make approximations and, of course, we try our best to simulate them with classical computers, but for some problems, we have to approximate to a degree that the answer we get out is just too fuzzy or is not meaningful."

\begin{figure*}[t!]
    \centering
    \includegraphics[width=.97\textwidth]{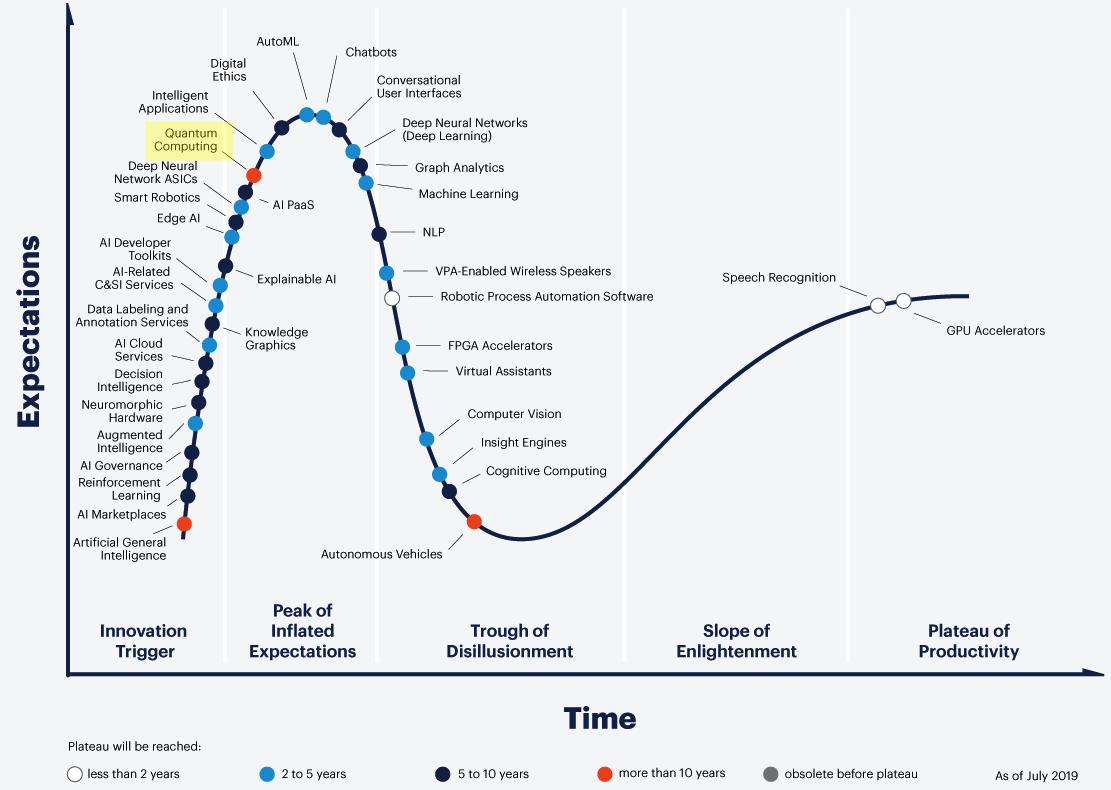}
    \caption*{The 2019 Gartner Hype Cycle for Artificial Intelligence, with quantum computing highlighted in yellow. Credit: Gartner}
    
\vspace{-.15in}
\end{figure*}

Professor Oliver provided a few examples of applications he is especially interested in and believes can be addressed with quantum computers. The first is the prediction of new materials that have certain functionalities, such as topological materials. In particular, he is hopeful that we could find a way to build materials that superconduct at room temperature. In the domain of quantum chemistry, Oliver is particularly interested in theoretical work performed at Microsoft with respect to the nitrogen fixation process. Currently, ammonia is manufactured using the Haber-Bosch process that was developed over 100 years ago, during WWI. Nitrogen fixation is a necessary step to create the fertilizer required to feed an ever-growing human population. However, it also requires extremely high pressure and temperature, contributing to an estimated 1-2\% of worldwide energy usage. We do not currently know how to improve the Haber Bosch process, because we do not understand how the main enzyme in the process catalyzes ammonia. Professor Oliver pointed out, however, that bacteria have an enzyme that serves as a catalyst for nitrogen fixation using the small amounts of energy provided by their metabolism. Thus, he believes there is a more efficient way we could go about this process and hopes that we can find it, using quantum computers to understand the chemical catalysis process. 


\vspace{-.05in}
\subsection*{Are superconducting qubits the quantum transistor?}

In order to describe the current state of quantum computing technologies, Professor Oliver felt it was necessary to briefly recount the history of classical computation. We began with the Babbage engine, a mechanical engine developed in the 1800s. This device worked in principle but did not scale well in practice, because it generated excessive amounts of friction. In 1906, the vacuum tube was invented and used for radio transmitters and receivers. 40 years later, the first vacuum-tube-based computers were developed for use in WWII. In 1947 the bipolar junction transistor was invented. Although we started making computers out of these discrete transistors in the 1950s (we did that here at MIT!), the memory was very different (it used magnetic core memory), it was not integrated, and there were a lot of solder joints, which had to be made by hand by some poor MIT graduate student or staff researcher. It was not until the 1970s that Intel started making integrated chips with thousands of transistors (roughly 4000). Then it would be another 20-30 years before Intel released the Pentium model and broke the million-transistor barrier. Another 20 years later we would go on to develop GPUs and multi-core processors, recently breaking the billion-transistor barrier. In summary, classical electronic computing spanned many different technologies, over more than 100 years. 

Quantum computing, on the other hand, started with Richard Feynman's suggestion, in the early 80s, that you need quantum to solve quantum. 15 years later, Peter Shor developed Shor's algorithm and, with some colleagues, created the CSS error correcting codes theory. Around the same time, Professor Eddy Farhi, Jeffery Goldstone, and colleagues came up with adiabatic quantum computing and Hidetoshi Nishimori came up with quantum annealing. In the roughly 20 years since then, we are now making chips with 10-50 qubits. So, Oliver believes there is still a long way to go. He thinks ``superconducting qubits are great, because they work and they work today, so we can do a lot with them. And they will push the field forward and help us learn a lot about all the other things that we need to learn how to do to make the technology."  However, in the long-term, Professor Oliver is not sure where superconducting qubits will stand in the history of quantum computation. ``Are superconducting qubits the vacuum tube? Are they the bipolar junction transistor? Are they complementary oxide MOSFETs? I have no idea of what it is! It may be that it goes in the history books in 5 or 10 years as some other technology takes over. I don't know. So, my perspective is that I enjoy quantum information science technology and trying to do quantum computing. I'm currently using superconducting qubits and I'll push it as far as it goes. After that maybe we'll have to jump onto another ship. Maybe we will get lucky and it's the transistor, but maybe we're unlucky and it's the vacuum tube. We'll see."

\subsection*{Thoughts on media hype currently surrounding quantum computation?}

Professor Oliver believes that there is a lot of hype right now in the field. He notes this is helpful in the sense that it brings awareness of the technology to the general public. This also helps in convincing funding agencies to give money to research in the field, as demonstrated with the recent approval of the National Quantum Initiative. While Oliver acknowledges that there are these positive sides to hype, there are also negatives. In particular, he worries that the expectations become too unrealistic and that the time frames over which these wonderful promises are going to be realized gets shorter and shorter. At this point in the interview, Professor Oliver pulled out his computer, to figure out where exactly quantum computation falls on the Gartner Hype Cycle. He believes that the technology is currently in the upper end of  the ``Innovation Trigger". Things are starting to work and promises are being made. However, if the hype cycle is correct, the field will reach the “Peak of Inflated Expectations” and then fall into the “Trough of Disillusionment”.

Professor Oliver expressed concern that the field is currently overhyped. He believes that the amount of time to realize promises made is longer than many are willing to acknowledge and hopes that the ``Trough of Disillusionment" which may follow is ``not too deep." Oliver thinks that coordinated government funding – such as the National Quantum Initiative -- can help with making it through this period. However, he feels that it is also important to continue working towards the goals that would make quantum computers profitable, such as finding a NISQ algorithm. ``It is not the holy grail, but can potentially solve some useful problem that people care about. Companies can then sell it and generate revenue.  That will help us find our way through any trough. Either that or we have to go fast enough that we jump the Snake River Canyon and go flying to the `Plateau of Productivity'...I do not think it works that way though. It would be cool if it did." 

\subsection*{What is Lincoln Lab's involvement in quantum computing research happening at MIT?}

\begin{figure*}[t!]
    \centering
    \includegraphics[width=\textwidth]{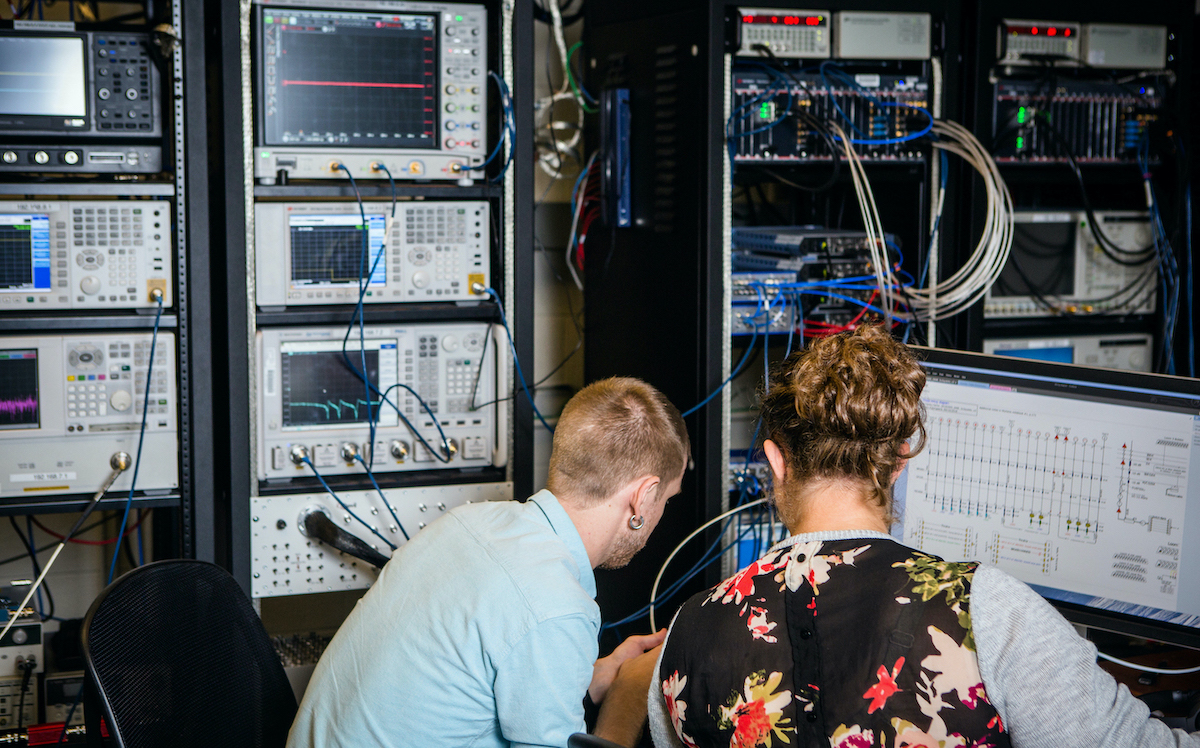}
    \caption*{Researchers, of the MIT Engineering Quantum Systems group and Lincoln Labs, setting up racks to control and measure
superconducting qubits. Credit: Nathan Fiske}
\vspace{-.15in}
\end{figure*}

Professor Oliver began working at MIT through Lincoln Laboratory, where he is currently a Laboratory Fellow. From the beginning, he collaborated closely with Professor Terry Orlando on the MIT campus side. He always felt that, for superconducting technology to work out, it would need both the exploratory research of an academic environment coupled with the directed research environment of a national lab. This has largely played out. The group that Oliver leads at Lincoln Laboratory performs research in areas related to developing advanced, reproducible quantum systems. It works on projects leading to, for example, higher circuit yield, by carefully studying the materials, taking a statistical approach to fabrication, addressing microwave package design, and building testbeds. Meanwhile, the Engineering Quantum Systems group, his group at the MIT campus,  does much more exploratory research. Professor Oliver says that this this type of exploratory research can cover more territory, be much faster, and quickly demonstrate new concepts.

He believes there are unique benefits of working at each of the two locations and aims to make it feel like one cohesive group. ``It's not us and them. It's all of us at MIT, the Institute. It's a one-team, two-locations kind of mentality. And that has worked very well." Both groups have grown significantly over the past few years. The broader quantum information group at Lincoln started over a decade ago with just a handful of people and now has over 100.  The campus group similarly has grown to over 20 postdocs, graduate students, and UROPs.

In fact, this Lincoln-MIT model has worked so well that the heads of the MIT Research Laboratory for Electronics and Lincoln Laboratory have initiated the RLE-Lincoln Center for Quantum Engineering, which aims to expand the Lincoln-MIT model to research in other quantum technologies, such as trapped ions. Among the Center's many goals, one is to foster greater collaboration between Lincoln Labs and campus. If it goes well in the quantum domain, Oliver believes that the approach may be expanded to other disciplines at MIT, such as aeroastro and biology. He believes that the RLE is particularly conducive to this type of model because it already is very interdisciplinary and crosses so many departments.

\subsection*{Thoughts on corporate involvement?}

Professor Oliver believes it is very good that start-ups and blue-chips are investing in quantum. While this brings marketing and hype, he believes that these companies are helping push the field forward. In reflecting on the history of corporate involvement in classical computing, Oliver believes it makes sense to be involved, even at this early stage. He says that computing companies that were successful in the 70s and 80s were the companies (or bought the companies) that were already working on the problem in the 40-60s. ``If you're not in the game, in some meaningful way, it's unlikely that you will be a leader in the future. You have to be in the game and you have to be pushing it forward, even though at any given time you might feel like we are far from there yet. But, if you wait for somebody else to do it, then they will have a leg up. And this is hard. It is not the case necessarily that you can just buy it when someone else develops it." Even from a national perspective, Oliver believes we need to be wary of letting others develop the technology and then buying it. Corporate involvement means progress is being made and it is only a question of how long. ``If you're saying too short time scales, then it's hype. If you're saying too long time scales, that it will never come, then you're a wet blanket. And if you're somewhere in between, then you're probably realistic." Furthermore, Oliver believes that corporate involvement is particularly beneficial for MIT, because not all graduate students will end up in academia, and many do not want to. Corporate development of quantum technologies ensures that the quantum workforce currently being trained at MIT will have a place to work in industry.

As part of the previously discussed quantum engineering initiative, there is an industrial consortium component, which aims build an ecosystem and educate the existing workforce. Both Professor Oliver and Professor Chuang led the development of an MIT xPro online course, along with Professor Harrow and Professor Shor, which targets mid-career professionals in industry and government. This 16-week course is geared towards providing a basic introduction to quantum computing, quantum algorithms, and the main challenges facing the technology. More than 1500 learners have taken the course in the past 18 months, far exceeding expectations, and it will continue to be offered for years to come. According to Oliver, ``we need to figure out what this quantum engineering discipline means. We need to understand what the foundations are of this new engineering discipline and somebody has to write the textbooks. We have to develop the courses and train students and professionals alike. Quantum engineering is a bridge between the scientific and engineering disciplines, and we will need to pivot all those fields: physics, math, computer architecture, analog/digital design, control theory, chemistry, biology, etc. All of these disciplines have something to bring to quantum computing. And you know, what a better place to do it than MIT?!"

\subsection*{Prof. Oliver's final thoughts.}

In wrapping up our interview with Professor Oliver, we asked if there was anything that should be emphasized in the article. He felt that one of the main takeaways should be that ``despite the hype, MIT has really been there from the beginning and pioneered a lot of work in this area. Almost every [quantum] algorithm that we know of today is connected to a person at MIT in some way." He gave specific examples of Professors Peter Shor, Aram Harrow, Ed Farhi, and Seth Lloyd. While these faculty may not have developed the algorithms they are famous for at MIT, they have all been MIT faculty at some point or are currently faculty. In addition to being a theory powerhouse, Oliver believes that MIT continues to excel in experimental work. Although his group focuses on superconducting qubits, he highlighted research on trapped ions, quantum optics, quantum communications, and quantum sensing. Professor Oliver believes that ``it is more than just quantum computing. It is quantum information science and quantum engineering, and that includes sensing, communication, networking, and computing."

\newpage

\section*{Professor Dirk Englund – \\Quantum Photonics}

Dirk Englund is an Associate Professor of Electrical Engineering and Computer Science at MIT and the Principal Investigator of the Quantum Photonics Laboratory. His research interests include quantum optics, precision measurement, and nanophotonics.

\subsection*{How would Prof. Englund explain his research to a non-expert?}
According to Professor Englund, information is always encoded in some physical system, whether that be a small magnet or a charge. As information storage is shrunk to smaller and smaller length scales, the regime of information manipulation and storage becomes very small-scale. At this small scale, nature behaves according to the laws of quantum mechanics, which is quite different from how we see nature in our everyday lives. This in turn puts a limit on how much we can shrink devices. After 60 years of semiconductor scaling, we are now finally hitting those length scales. At tiny length scales, it is no longer possible to confine electrons. Through a phenomenon known as \textit{\color{DarkRed}quantum tunneling}, electrons can hop from one device to another and tunnel through barriers they are not supposed to. According to Englund, ``if you are a pessimist, you might say this is the end of the scaling laws. We can no longer improve the performance of computers and so on that have changed our lives. But if you're an optimist you can say, maybe it's an entirely new chapter, where you look at the glass half-full rather than half-empty. And that chapter is the quantum information era, where rather than bemoaning the strange properties of nature at the atomic scale, you make use of them." 

Through his own research, Professor Englund hopes to explore new opportunities that emerge when encoding information in quantum physical systems, particularly those based on photonics. ``This could lead to computers and other kinds of information processing devices, as well as sensing devices that can do things that we do not know would be possible in a world governed by classical mechanics." In the 1950s, the first transistors were made out of silicon. It took a couple decades until these transistors were refined to the single transistor level and then could be replicated and multiplexed. Englund believes that, today, we are in the quantum information era at a level in which we can reliably create single ``quantum transistors." However, they are still bulky and they are hard to manage. Through quantum photonics, researchers are learning how to put many of these quantum “transistors” on devices and how to scale them up.

\subsection*{How did Prof. Englund get into quantum photonics research?}

As an undergraduate physics major at Caltech, Englund was very interested in both fundamental physics and applications research. He worked as an undergraduate researcher in the group of Professor Hideo Mabuchi. The work focused on controlling quantum information in atoms. Professor Englund provided a vivid example in which the group used electron and nuclear spins in atoms to store information locally and then transmitted the information optically, via photons. This is analogous to a computer, in which single atoms were used to store information. However, transmission occurred through photons, rather than electrical currents, because photons are better decoupled from the environment. 

\begin{figure}[t!]
    \centering
    \includegraphics[width=.9\columnwidth]{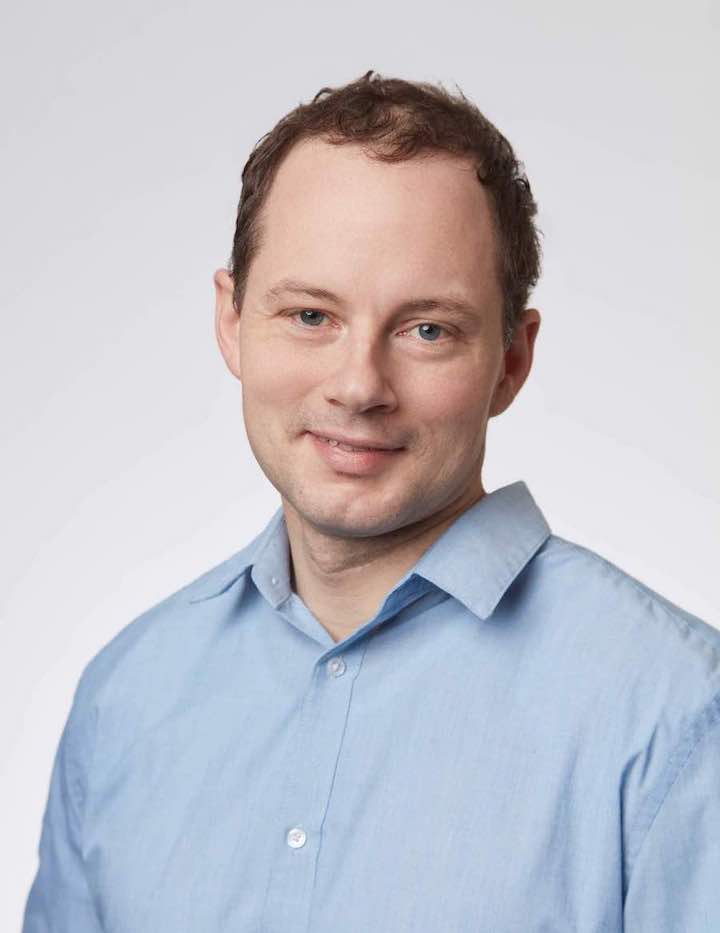}
    \caption*{Professor Dirk Englund, PI of the MIT Quantum Photonics Laboratory research group. Credit: MIT RLE}
    \vspace{-.15in}
\end{figure}

Englund wanted to see if he could make these devices more practical by implementing them in solid state systems. Rather than using atoms suspended in space, he wanted to use artificial atoms, or things that behave like atoms (as described by Prof. Oliver), in solids. In particular, Englund looked to \textit{\color{DarkRed}quantum dots}, or nanoscale pieces of solid semiconductor that have optical and electronic properties very similar to single atoms. One key advantage of quantum dots is that they are stuck in a solid, meaning they do not need to be trapped in a vacuum. Englund believed they would thus be easier to manage.

He pursued this idea through his PhD work at Stanford. He focused specifically on controlling individual artificial atoms and interfacing individual artificial atoms with individual photons. ``That was the start of quantum for me. Now we are at a point where we can control that interaction very well. We have pretty good memory in these artificial atoms, in the form of electron spins and nuclear spins. And we are actually getting to that point of scaling." Professor Englund considers his graduate research as part of a larger effort towards creating the quantum “vacuum tube.” As a professor at MIT, he is now thinking about how these systems can scale and how they can be used to create integrated circuits.

\subsection*{How does quantum photonics compare to other hardware approaches?}

Professor Englund believes that photonics will play a major role in any quantum computing technology, at some level. In fact, in some it is the central piece. Several companies are currently trying to build quantum computers entirely out of photons. However, even if this is not the case, photons are integral to most hybrid systems. Modular quantum computers will store information in the form of internal states of matter, such as atoms or trapped ions. A leading scheme to connect large numbers of these atoms or trapped ions, is via photons. Englund claims that quantum photonics is central to this effort, as a way to ``make the wires that connect all the qubits." Finally, in the case of superconducting circuits in particular, he argues that they are simply a means of doing photonics at the microwave level -- a microwave optical quantum computing approach. In fact, there are many ongoing efforts around the world in transducing quantum bits from superconducting microwave photons to optical photons, using special interfaces. This would allow researchers to connect quantum computers together over distance, such as through a city or data center. As of right now, photons are the only viable means of connecting quantum computers at a distance because they are very weak coupling to the environment. Thus, Englund argues that regardless of whether it is photonics themselves, trapped ions, or superconducting qubits that pan out in the long-term race to build a quantum computer, photonics will play an integral role nonetheless.

\subsection*{What is the trajectory of quantum computation over the next few decades?}

In 2016, MIT Professors Dirk Englund and Seth Lloyd organized a government sponsored workshop titled the \textit{``Future Directions of Quantum Information Processing: A Workshop on the Emerging Science and Technology of Quantum Computation, Communication, and Measurement"}~\cite{future}. Several of the top scientists from the US and abroad gathered in Arlington, Virginia, to identify challenges and opportunities in quantum information processing for the following 5, 10, and 20 years. Englund summarized the key takeaways of the workshop for us.

By the end of the 5 year timeline, it was anticipated that we will be in the, previously defined, NISQ Era and that we will have quantum systems too large for any classical computer to predict the behavior of. Although he believes they will have a big impact on computer science related complexity arguments, Englund's primary question is whether these systems will be useful for something practical. He cited a few particularly promising examples of current applications research that he was excited about. In terms of simulating other quantum systems, like molecules, there is potential for better first principle structure design in materials development. In terms of  optimization problems, there is potential to optimize logistics chains and other related. And finally, in terms of specialized algorithms, there are versions of algorithms, like Shor's, that have been proposed which could potentially run on NISQ quantum devices. Alongside all this progress in applications, Englund looks forward to progress in benchmarking studies, “which pitch quantum computers that are going to be imperfect and have limited numbers of qubits and gates against classical computers. I think that we can do this at all is amazing, because classical computers have been developed for much longer and they have way way way more money being poured into them. [The fact] that there is benchmarking on certain problems is really interesting and exciting.” 

Looking out to the 10 or 15 year time frame, Englund hopes that the noisy aspect of these intermediate quantum computers will be fixable using error correction. With this error correction, he claims we will have more confidence that error-corrected general-purpose quantum computers will have useful applications. With regards to Shor's algorithm, he believes that there is a reasonable chance that before 2030 there will be a quantum computer which can break RSA encryption at a level not doable with classical computers, ``let's say about 1800-bit encryption." Englund says that even if there is only a 5\% chance that in the next 10 years a quantum computer will break RSA, we should be very worried. ``5\% is a lot. Especially if you don't know if somebody has that computer."

\begin{figure}[b!]
    \centering
    \includegraphics[width=\columnwidth]{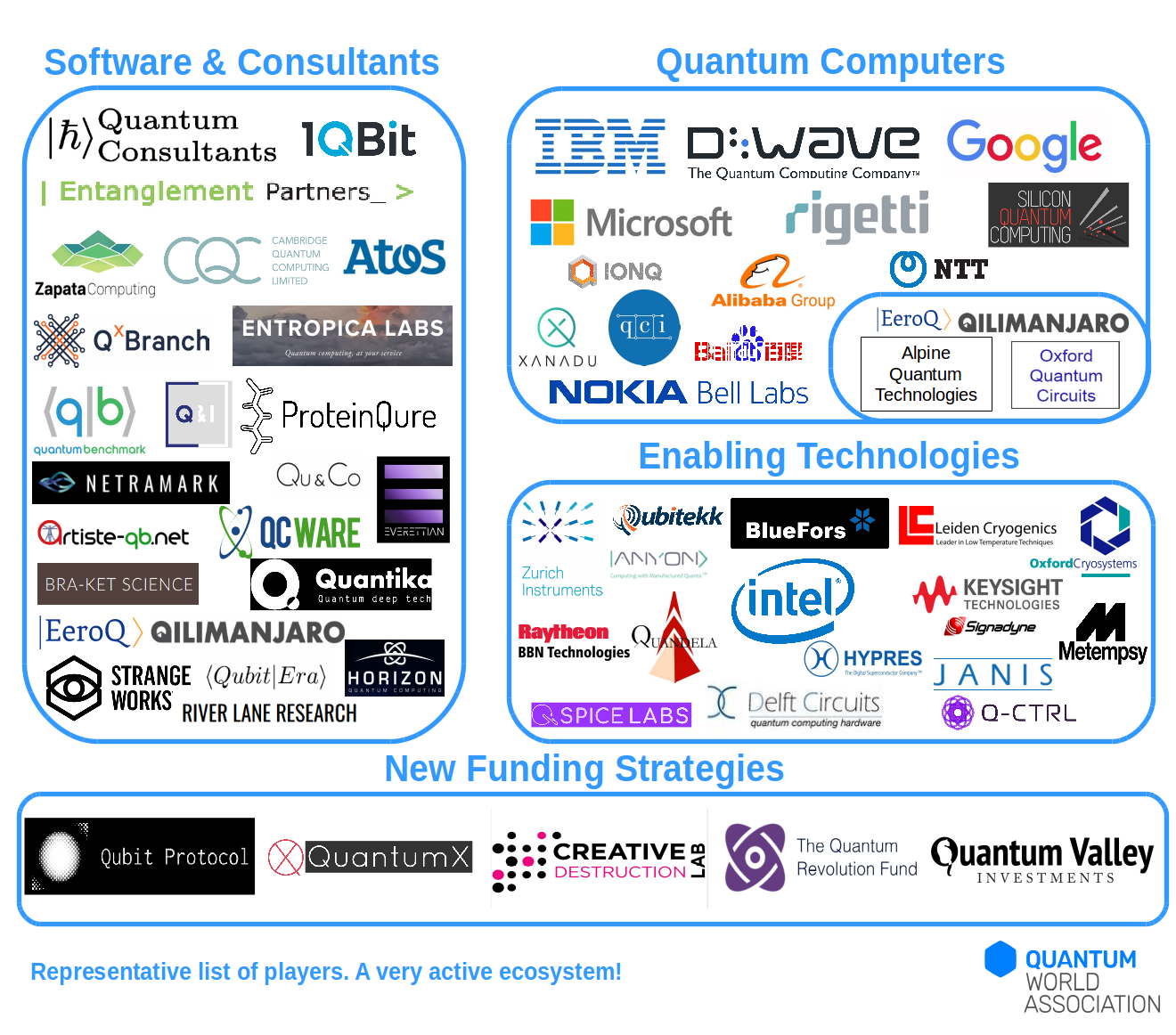}
    \caption*{Companies and startups involved in the 2018 quantum computing ecosystem. Credit: Quantum World Association}
\end{figure}

\subsection*{Thoughts on current media hype? Will there be a quantum winter?}

No matter what, Professor Englund believes that a lot of fantastic science that will come out of all the current research into quantum computing, both in terms of theory and experiment. With regards to commercial efforts in particular, he thinks it is fantastic that there is commercial activity. This allows systems to scale up much more quickly. That being said, Professor Englund believes it is still unknown what the most successful systems will be. ``Some approaches will probably fold and some will continue to prosper. And then perhaps one can assign hype later, but it's tough to predict upfront what's going to be the best approach." Overall, however, he is optimistic that there will not be a general downturn on hype. ``I think there will be some approaches that perhaps go away and some that will come up and some that will continue. That could make the field stronger as a result of it." He does not foresee a winter as dramatic as those experienced by the artificial intelligence community in the late twentieth century.

In fact, Englund is confident there will continue to be big advances in experimental and theory efforts. ``Some systems, once you investigate them for long enough, yeah maybe they won't scale. But that's ok. There are other systems that I think are going to go forward. I think there's several systems that I'm actually quite optimistic about. But it's not around the corner. I think some people in the media like to give the wrong perception that a general purpose quantum computer is a couple years away. It's probably still a decade away for a general purpose error-corrected computer. Maybe longer than that. But there's a lot of interesting problems in-between, interesting questions that are going to be answered. So, in my view, it is going to keep the field exciting, even before you have that ultimate goal of a general purpose error-corrected quantum computer. Which is fortunate for us, right, there are intermediate scale systems you can use before you have a full blown computer. That's super important."

\begin{figure}[t!]
    \centering
    \includegraphics[width=.9\columnwidth]{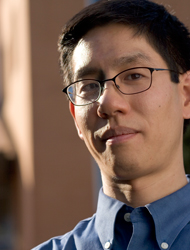}
    \caption*{Professor Isaac Chuang, PI of the MIT Quanta research
group. Credit: MIT Physics}
\end{figure}

\newpage

\section*{Professor Isaac Chuang – \\NMR \& Trapped Ions}

Isaac Chuang is a Professor of Physics and Electrical Engineering, as well as the Senior Associate Dean of Digital Learning at MIT. He is the Principal Investigator of the MIT Quanta research group and a co-author of ``Quantum Computation and Quantum Information," the primary textbook in the field.

\subsection*{How would Prof. Chuang explain his research to a non-expert?}

To a non-academic, Professor Chuang would describe his research as an attempt to build computers out of single atoms and single electrons. He says this is ``cool" because these atoms obey special laws of physics, that can be used to solve problems that cannot otherwise be solved. However, if his middle-school daughter were to ask if she could use one, he would have to say ``no sorry, not quite yet."

For someone with professional knowledge of science, Chuang says he is ``trying to build a computer that can solve certain kinds of mathematical problems much faster than is possible with normal classical laws of physics, by using Schrodinger's equation." He claims that this is hard because quantum properties quickly disappear as we build larger systems. However, it is now believed that we can in fact keep those quantum properties intact, even while we scale our systems. In terms of exciting use cases of the technology, he cited the two major examples of Shor's and Grover's algorithms.

\subsection*{How did Prof. Chuang get into quantum computing research?}

According to Professor Chuang, there were two defining moments  that lead him to quantum computing research. The first was rather early on, roughly 30 years ago as undergraduate at MIT, when he realized that he loved both computers and physics.  ``I wanted to do something that would let me play with both of these ideas. And it was a little sad for me to realize that I had to choose a major, which is either 6 or 8. I really wanted to do both."  He started in 6 (EECS), but felt ``a little bit bored and decided to learn about why things were the way they were." Eventually he decided to double major in 8 (Physics).

Chuang claims that he only became interested in quantum computing per se when he read about Richard Feynman. He recounted digging through the MIT Physics Library to find Feynman's undergraduate thesis and reading any Feynman text he could lay his hands on.  When Chuang started grad school, he decided that he did not want to do anything that was popular at the time. Instead, he ``wanted to set out on [his] own and show that some of Richard Feynman's ideas would be feasible." He managed to convince a new professor, who had just joined the faculty, to let him ``go off and wander around doing this, even though nobody was doing it." He believes there were only 6 people doing that kind of research in the world at the time. They ``were all corresponding and saying this is a cool idea and thing to do, because there was no popularity in the concept of a quantum computer back then. There was Richard Feynman's 1985 article in Optics News and all of us who knew about it would get excited about it, but that was it." Chuang wrote an article called \textit{``How to Build a Simple Quantum Computer"}~\cite{chuang1995simple} that was accepted to Phys Rev A and then Shor's algorithm hit the news. ``One of my friends faxed me Peter Shor's article. And this is a preprint. No publication was able to publish as fast as the fax machines were going. I read it, I understood it, and I went around explaining to everyone I could possibly talk to. And I sat there going, I've gotta be able to realize this! After all, I had been one of the people working on the subject for fun. So then I set out to build it. And those were the early days. So, there was luck, but there was also this sense that I wanted to do something that wasn't possible. And I hope that's true of many of the undergrads today too."

\subsection*{What problems interest Prof. Chuang today?}

Given Professor Chuang's expressed interest in unpopular and unconventional topics, we asked him, if he were a student today at MIT, whether quantum computing would be too popular of a field for him. He replied that while quantum computing is very popular in physics, almost no engineer knows how to deal with quantum computing. Thus, he believes that quantum engineering, or the offshoot effort to do so, is what will grow big in upcoming decades. With this will come tools to engineer quantum systems into larger systems and an opportunity to look at quantum physics in new ways. Although he believes this will be interesting for both physics and engineering, he claims that as a physics undergrad today, he probably would not work on it. Instead, he would look at the physics of computation underlying it. ``How do physical systems compute and why? How do you take ideas from computer science and map them onto physical systems to understand them better?"

In particular, Chuang said that he would be interested in looking more deeply at the intersection of machine learning and physics. ``Machine learning seems to be effective on problems which relate to the real world, physical systems to spoken language to visual things in the 3-dimensional real space. Why? Why is machine learning so good at interpreting things in the real world. It's actually pretty poor at doing things which are algorithmically complex and that are abstract rather than real. So there's something interesting about how computers can extract information about the real world. And I think that understanding of machine learning can provide us with a lot of insight about physics. And so, I'm quite excited about that direction of machine learning affecting physics and also the fact that we may be able to realize these machine learning algorithms with physical systems...The fact that we might be able to do so much better and reduce the energy requirements of computation is very fascinating. It's a challenge, and it's something that you wouldn't be able to do as a machine learning person alone. You need to know the physics side of it. And that's something inspiring for today's generation, if I were in today's generation."

\subsection*{What were the expectations when Prof. Chuang first joined the field of quantum computation? }

\begin{figure}[b!]
    \centering
    \includegraphics[width=.77\columnwidth]{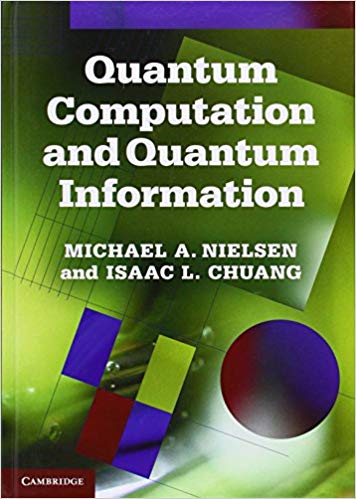}
    \caption*{One of the first and most prominent textbooks on quantum information, co-authored by Prof. Chuang. (Published 2000)~\cite{nielsen2002quantum}}
\end{figure}

Professor Chuang says that when he joined the field, it did not really exist. This meant that the few people working on the problem, at the time, set their own expectations and ``they were a little crazy." In fact, he says that most people at the time expected quantum computing to be an utter failure. Even after Shor's algorithm came out and quite a few people were trying to rapidly realize the technology, they worried that quantum computing would follow a long history of alternate algorithm paths, like analog computation, which failed largely because of noise. However, two advances came shortly thereafter, also by Peter Shor, which provided a great deal of hope for the field. The first was a seminal paper, which proved that there could be error correction for quantum systems. According to Chuang, ``this is wild because a quantum system seems to be continuous valued and yet you can correct it." The second major breakthrough was a series of three seminal papers, which ``showed why you can build an arbitrarily sized large-scale quantum computer out of competing parts that fail with certain probability, as long as that probability was lower than some threshold." This essentially reworked VonNeumann's theory of fault-tolerant computation for vacuum tubes into something useable for qubits. Professor Chuang claims that ``if it were not for those papers [by Peter Shor], the whole field would have died." Furthermore, he believes Shor's ideas are why he would say that the expectations of the time were exceeded. ``The biggest result for the first 20 years of quantum computing is largely that quantum computers are real. They are not just a theoretical abstraction. They can be realized in the laboratory."

\subsection*{What is the current state of quantum computing technology?}

Professor Chuang believes we are solidly in the quantum ``Vacuum Tube Era." He bases this claim off a great deal of reading he has done on the “Vacuum Tube Era” for classical computing. ``If you go to the computer museum in Mountainview in California, you will see that companies actually built a large number of amazing vacuum tube computers, that did very sophisticated tasks. They were all made totally obsolete by the silicon transistor when it came along. But vacuum tubes actually went really far. And so, you might be tempted to think what we have today with quantum computers is already going far enough that you might call it a non-vacuum tube. However, they still fail, they fail with exceedingly high probability, and we don't know how to step them up besides using error correction codes." Chuang notes, however, that there are theorists dreaming up alternate implementations of quantum computers which may overcome this challenge. In particular, the Microsoft Quantum research team has been looking into a method called \textit{\color{DarkRed}topological} quantum computers. Although extremely challenging to realize, such qubits would fail with much lower probability. ``It is a very, very difficult route, but maybe something like that, someday, is something that I would call a quantum transistor."

\subsection*{What are the big challenges for quantum information?}

Professor Chuang believes that the biggest problem currently facing the field is that there are really only two types of quantum algorithms. The first is the sub-exponential speed-up provided by versions of Grover's algorithm. The second is  the exponential speedup provided by variants of Shor's algorithm, which all involve a Quantum Fourier Transform (QFT). There are several algorithms of this form, such as the Harrow-Hassidim-Lloyd algorithm, which gives exponential speedup for solving linear systems. However, Professor Chuang finds the fact that these algorithms all use the same structure as Shor's algorithm ``rather frightening and disappointing." He believes that there should be other quantum algorithms, yet we have been struggling for over 15 years to try and discover them. ``If we don't have insight into what other, different kinds of algorithms might exist and why they might be useful, then the field won't go terribly far."

Chuang also worries about the vast amount of speculation as to what might be feasible. He claims that people are coming up with all kinds of uses for quantum hardware, without knowing in principle how they ought to behave. To illustrate the point, he drew an analogy to machine learning in classical computing. ``It just does well. People don't know why it does well. You can't prove any bounds on it. People are throwing the same logic at quantum computers and building quantum variational autoencoders or other kinds of variational quantum algorithms, with no proofs. So, something has to progress along both of those lines." 

\subsection*{What is the trajectory of quantum hardware?}

With regards to quantum computing hardware, Professor Chuang believes the field will enter an engineering path, analogous to Moore's law. He specifically referenced the quantum analogy, called \textit{\color{DarkRed}Schoelkopf's Law}, which states that quantum decoherence will be delayed by a factor of 10 every 3 years in superconducting qubits. Before becoming a faculty at MIT, Chuang worked at IBM on superconducting technology. From his experience, Chuang believes that improvements in coherence times will eventually taper off. However, he is excited by the rapid growth in the number of qubits the field has seen over the past 2-3 years. He claims that, independently, the number of qubits or the coherence times themselves are not particularly good metrics. However, he believes IBM's \textit{\color{DarkRed}quantum volume}~\cite{bishop2017quantum}, is a good metric. This metric accounts for the total number of qubits times the total amount of coherence time available for the qubits of a device. Thus, it provides an idea of the volume of computation of a quantum circuit, estimating how much space and time are required to perform a computation. Professor Chuang believes that quantum volume ``might increase fairly rapidly," meaning that either the number of qubits or total coherent time will increase.

\begin{figure}[b!]
    \centering
    \includegraphics[width=\columnwidth]{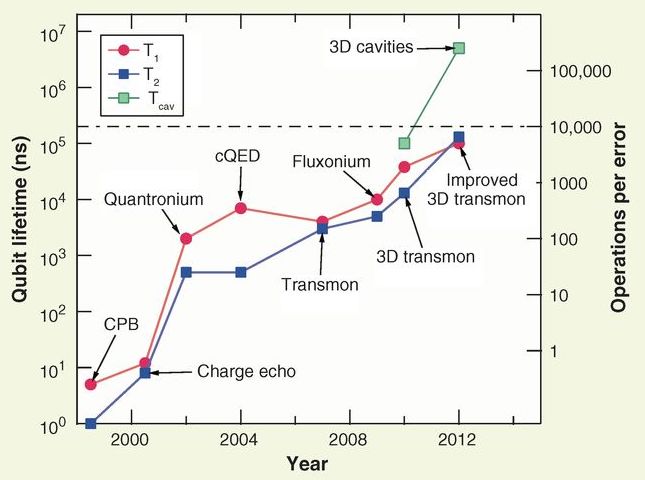}
    \caption*{An early figure indicating exponential improvement in qubit lifetime, supporting the proposal of Schoelkopf's Law.~\cite{devoret2013superconducting}}
\end{figure}

While he mostly referred to superconducting qubits in his previous response, we also asked Chuang for his thoughts on other prospective quantum computing modalities. He believes that each technology has its own special features and gave us a few concrete examples. According to Chuang, \textit{\color{DarkRed}trapped ions} have optical control and optical sensing abilities, meaning that they can integrate with long distance telecommunication systems very well. A superconducting qubit, on the other hand, has no optical interface, meaning it will not be a good communication system. However, the cycle time (or basic clock rate) for superconducting qubits is on the order of nanoseconds, which is far faster than the 1-100 microsecond times of trapped ions. ``Why is that relevant? Some people have talked about using quantum computers for high-frequency trading and finance. If you have clock times in the microseconds, there is no way it will be useful in high-frequency trading, whereas superconducting qubits might. On the other hand, the superconducting qubits' coherence time is on the order of milliseconds, so they lose their memory very fast. The trapped ion qubits have a coherence time demonstrated to be on the order of tens of seconds. So you might think about using both of them." While Chuang noted that many researchers are thinking about this specialization of quantum hardware, such as using different qubits for processing and for memory, he was much more interested in the question of what a quantum power supply should be. He argued that unlike a classical computer, it need to supply not just the source of electrons, but the source of pieces of computation. If fact, he believes that a quantum power supply should provide entangled states, which need to be routed to the right units and can be used to do things like convert from one code to another code.

Overall, however, Chuang made it clear that ``almost no technology continues on as first envisioned.” He believes that certain things will remain constant in quantum computing, such as “building our quantum systems on top of the best classical technologies." Superconducting qubits currently do that by making use of state-of-the-art photolithography techniques. However, trapped ion qubits are also beginning to do this by making their ion qubits out of lithographically patterned chips. Chuang feels these kinds of advances are necessary for the technology to develop into large systems. ``We can't custom build everything, we really have to build it on top of the best available technologies to get to large-scale systems."

\begin{figure}[b!]
    \centering
    \includegraphics[width=\columnwidth]{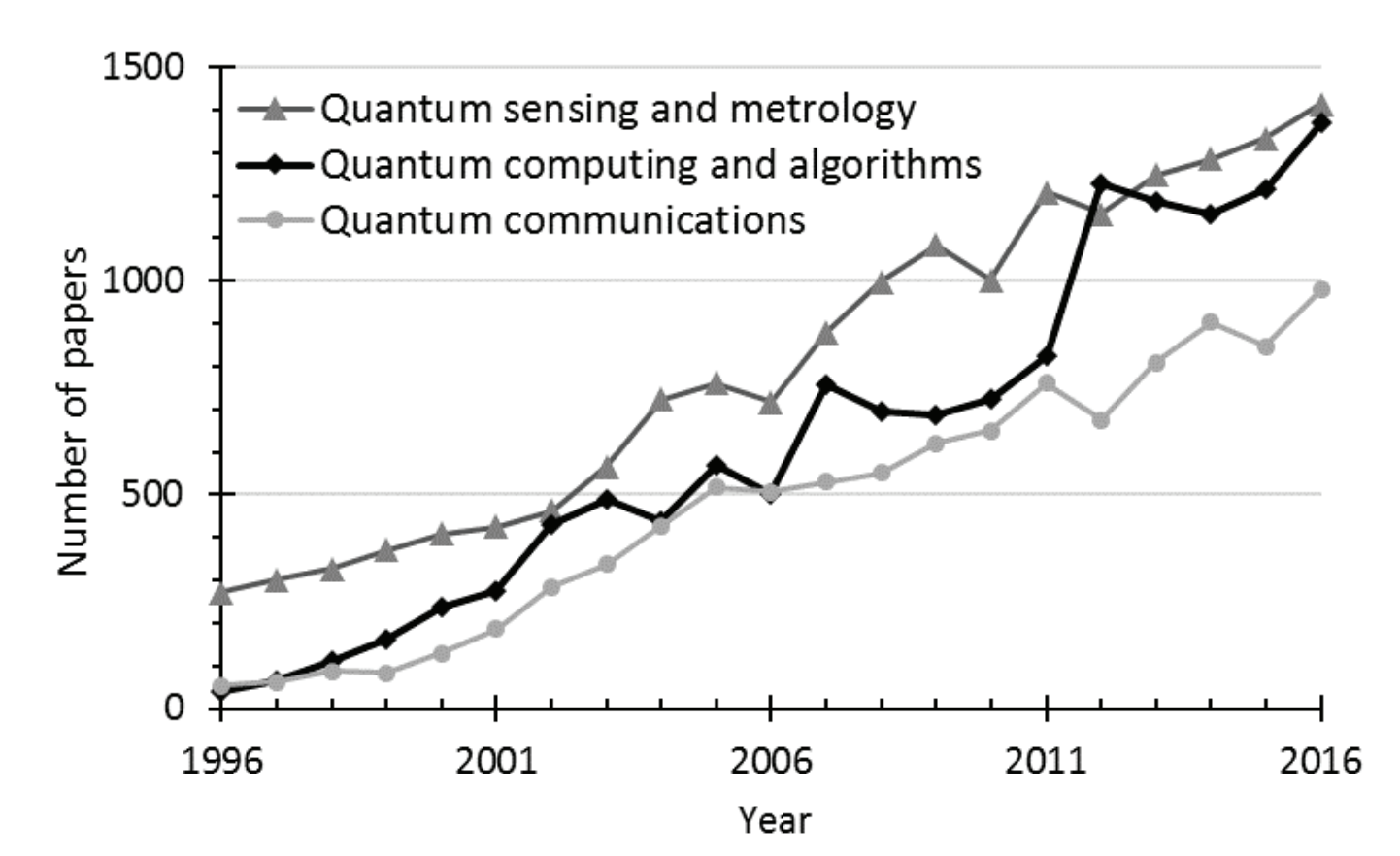}
    \caption*{The number of research papers published per year in quantum
computing and algorithms, quantum communications, and quantum sensing and
metrology.~\cite{future}}
\end{figure}

\subsection*{What kinds of problems will we use quantum computers to solve?}

Professor Chuang believes that we will use quantum computers for two different classes of problems. The first is Shor's algorithm applied to cryptography, for number factorization. Chuang is confident that the algorithm will work, in part because he demonstrated it in small-scale at IBM and again, over 10 years later, with trapped ions. He claims that the algorithm is “real,” but will only be useful when it can be used to factor hundreds of bits. In order to realize that, we will need hundreds of reliable, fault-tolerant qubits, each of which might require 7-49 more qubits. So, it currently seems that to run Shor's algorithm for the purpose of cracking RSA would require tens of thousands of physical qubits. Considering that we are at systems of less than 100 working qubits today, Professor Chuang thinks it will be roughly 5-10 more years until we can run Shor's algorithm in full scale. However, he is fairly confident we will reach tens of thousands of qubits in the next decade because ``engineers are getting good at [making qubits]."

In addition to cryptography, Chuang believes that a major application of quantum computers will be quantum chemistry. ``And that is interesting because it's kind of like machine learning. The natural thing you apply your algorithm to is what the natural world maps to your algorithm." Nowadays, we understand the structural rules behind these chemical systems very well. According to Chuang, while we have the physical and mathematical intuition developed to simulate large systems, the necessary computational power exceeds the limits of classical computation. Thus, we can make use of quantum systems. In the words of Chuang, ``this is Richard Feynman's idea evolved from the very start."

\subsection*{Thoughts on current media hype? Will there be a quantum winter?}

Professor Chuang believes that ``it is almost inevitable that this field will experience such waves, boom and bust." Although he is not sure of what a quantum winter would look like, he believes one is impending given the speed of modern communication and extent of simplifications made in the media. Chuang claims there are two competing timescales underway. The first is a slow timescale associated with the challenging problem of building a quantum computer. The second is a rapid timescale of news, communication, expectations, finance, and funding. ``A typical venture capitalist wants a return on investment in a 5-year timeframe. That's why people talk about 5, 10, 15 [years], because that's the structure of financial systems today. And if you can't yield some return on investments in 5 years, then you're not ready for that kind of investment. And yet there are people willing to take that risk and gamble on getting something in 5 years, today. So, that timescale of 5 or so years sets a natural cycle for winters and warm cycles. We will see what happens. It's very likely there will be some kind of bust and boom."

\subsection*{Thoughts on corporate involvement in quantum computing?}

To answer our final question, Professor Chuang wanted to take a step back. In looking at quantum technologies from a broader perspective, we must also consider the technologies of quantum sensing and quantum communication. According to Chuang, these two areas are also growing, but differentiate themselves from quantum computing in a very distinct way. ``In particular, they are yielding economic value, because you can sense biological systems, you can sense magnetic signals, better than you can classically in practice, in real systems. And so, there's value that those systems and quantum communication provides economically." For quantum computation, Chuang does not yet know where those economic benefits lie or when they will be demonstrated.

That being said, Professor Chuang greatly values corporate involvement with the technology. He claims it is, in fact, part of the ``natural lifecycle of invention and creativity." First, something is invented conceptually and demonstrated in the academic environment. However, if it is going to succeed, it usually transitions outside of academia into the next stage, whether that be governmental or corporate. This is necessary to show that the technology is actually useful to the economy, for solving a practical problem. And while, Chuang believes this cycle is very ``healthy," he notes that it has to be done just right, ``otherwise you get too many of these winters." He then proceeded to list several technologies, such as AI and optical computing, which did not get the timing right. “Physics is no stranger to these cycles as well. So, we don't know if the timing is right, but it's surely going to happen." 

\begin{figure}[t!]
    \centering
    \includegraphics[width=.9\columnwidth]{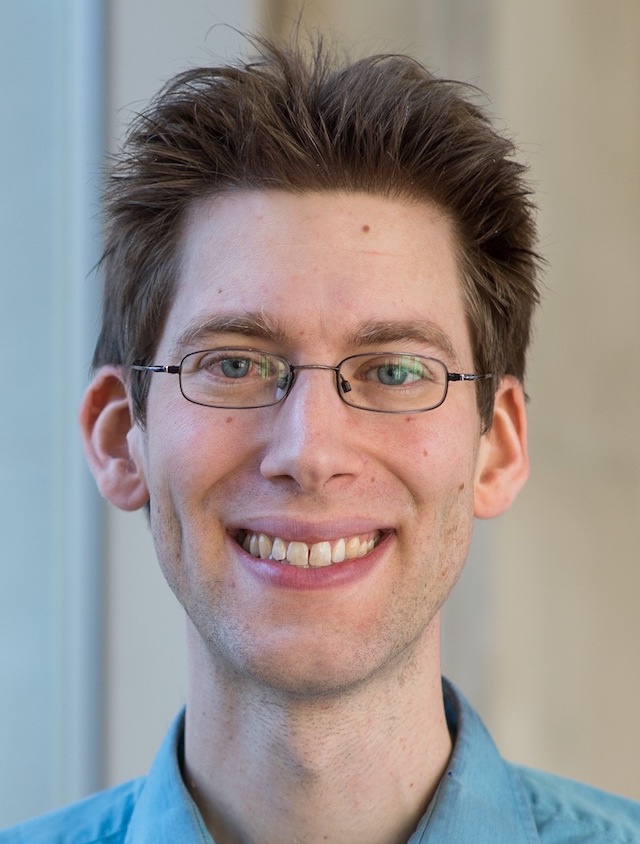}
    \caption*{Professor Aram
Harrow of the MIT Center for Theoretical Physics. Credit: Justin Knight}
\end{figure}

\newpage

\section*{Professor Aram Harrow -- \\Quantum Information}

Aram Harrow is an Associate Professor of Physics in the MIT Center for Theoretical Physics. His research interests include quantum algorithms, quantum information, quantum complexity theory, representation theory, and optimization.

\subsection*{How would Prof. Harrow explain his research to a non-expert?}

Generally speaking, Professor Harrow describes the goal of his research as ``figure[ing] out what to do with a quantum computer and quantum communication network, if we had them." Additionally, he aims to use ideas from quantum information to improve our understanding of physics and theoretical computer science. ``Quantum mechanics was originally seen as a theory of physics, about small particles and how they behave. Recently, we have really begun to appreciate that it has a lot of implications for information and that this has emerged as a field in its own right. How should we update things like information, proofs, computing, and communication in light of quantum mechanics? And so, if you were to talk about what I did broadly, you could say that I just look at theoretical aspects of what quantum mechanics means for information."

\subsection*{How did Prof. Harrow get into quantum information research?}

As an undergraduate at MIT, Harrow knew that he liked math, physics, and computer science. He also knew that he did not like choosing between things. So, “without a very good understanding of what the range of things in the world was,” quantum information seemed like a way to do all the things he was interested in. Harrow also cited a few, early career influences. This included a talk he attended by Professor John Preskill, a famous quantum information researcher from Caltech, at the weekly MIT Physics Colloquium. Additionally, Harrow did a UROP in Professor Neil Gershenfeld's lab, working on an early NMR quantum computing experiment. Overall, Harrow appreciated the idea that ``information is not the same as what your commonsense view of information might be and that you need physics to say what information is." In particular, quantum mechanics reveals that information, computing, communication, secrecy, privacy, and correlation are ``different and richer than we originally thought. I always thought that was really cool. I kind of liked that from the beginning about the field." 
	
Thus, Harrow got into quantum computing late in undergrad. He stayed at MIT for a PhD in the Physics department, knowing that his specialty was indeed quantum computation. Postdoctorate, he lectured in Mathematics and Computer Science at Bristol University for 5 years and served as a Research Assistant Professor of Computer Science at the University of Washington for 2 years, before returning to MIT.

\subsection*{What were the expectations when Prof. Harrow first joined the field of quantum information?}
Professor Harrow notes that the field was much smaller when he first started his UROP in the summer of 2000. He feels that at the time, ``it was much less clear it would work. The noise rates were intimidatingly large and progress to improve them was very slow. Actually, the rate at which noise was being reduced was probably the same rate it is today. It's just that we've seen 20 years of steady progress, which has made everybody more optimistic.” However, as the experiments got better, interest in theory increased as well. “So, what shifted was that the noise just steadily got lower and lower. The experiments just got better and better. As a result, there was more interest from industry and from government funding agencies. And very recently, a lot more academic jobs. And so that's meant that there has been more interest in things like theory."

Professor Harrow admits luck in terms of the time he joined the field. He believes it has grown significantly since then. And with this growth, he believes that quantum information has ``emerged as a topic in its own right…It used to be that quantum computing conferences were full of people who were really physicists or computer scientists or mathematicians. In that, they got their training in some discipline and then came into the field. And now there are more and more people like me, where from the very beginning we were trained in quantum computing.” Harrow notes that this transition has both pros and cons. It has been beneficial in the sense that everyone is ``on the same page." However, he finds the loss of ``diversity of intellectual tradition" to be a shame. 

\subsection*{What does Prof. Harrow consider to be his most influential work?}
Harrow's most cited paper is the \textit{``Quantum Algorithm for Linear Systems of Equations"}~\cite{harrow2009quantum}, developed with Professors Hassidim and Lloyd. He believes this work has been very influential in leading to more quantum algorithms work along the same direction. In particular, Harrow thinks the paper helped researchers realize that they can perform operations on quantum computers which are a bit non-unitary, challenging the previous view that quantum computers are solely useful for unitary transformations. He  feels that this has ``helped people appreciate the flexibility of quantum computers a lot more and has led to people coming up with algorithms for solving differential equations and various machine learning tasks." However, Professor Harrow finds the whole area to still be a ``a little bit incomplete," solving only part of the challenge of building a quantum algorithm. There are aspects, such as choosing a useful input to the linear systems algorithm and what to do with the output, that still require more research. ``It's a bit like the way Shor's algorithm for factoring made use of the subroutine of the Quantum Fourier Transform. Peter Shor had to do a lot of work beyond the Quantum Fourier Transform to turn that into an algorithm for factoring. So we've, with this linear systems algorithm, made a nice building block, but more pieces are needed to connect it to applications. That's definitely been my most influential work."

Professor Harrow noted, however, that there are number of topics that he has really enjoyed working on, which he feels the overall community is less interested in. One topic area of particular interest for Harrow has been ``a cluster of questions that connect optimization problems, even on classical computers, with entanglement in quantum systems." In work done with Dr. Ashley Montanaro of the University of Bristol, Professor Harrow developed a procedure to test whether a pure multi-party quantum state is entangled or not. In his work, he demonstrated, ``in a not very obvious way," that it is hard to distinguish classical correlation from quantum correlation. ``It's computationally hard in the sense that I might have a complete description of the quantum state, but if I feed that into a computer and ask it to tell whether it has quantum entanglement or just classical correlation, that's hard to figure out. And then, this in turn is connected to a bunch of other computational problems, a lot of things involving tensors, which are like these higher dimensional generalizations of matrices. A lot of problems about tensors, that people didn't know were hard, we were able to show were hard by making this connection to entanglement." Although most of the work centered around claims on hardness, Harrow notes that some of this work also suggested new algorithms, ``either to approximately figure out whether a state is entangled or to approximately solve these tensor problems." However, the most ``fun" aspect of the work for Harrow was ``the diverse fields that got connected to [it]. I found myself collaborating with theoretical computer scientists and making unexpected connections to work on this question. That's been something I've enjoyed which has not been as widely influential, but that I really like."

\subsection*{When does Prof. Harrow expect his algorithms to run on a physical quantum computer?}

Professor Harrow argues that ``things like [his] linear systems algorithm are already being implemented, but just in toy ways." Although demonstrations on small-scale quantum computers cannot exceed the limits of a classical computer, he feels that these demonstrations are a good ``proof of principle." That being said, Harrow is not sure when his algorithms will be implemented in a useful way. ``It's hard to say, it just depends on the overall rate of experimental progress. And I probably don't have any unique insights into this. Maybe 5-20 years." However, Harrow is hopeful that it will be sometime in the near future, ``no thanks to [his] own work." He believes that the current rate of experimental progress and significant resources being poured into experimental efforts will allow these quantum systems to scale up rapidly. Professor Harrow notes, however, that his own algorithms are not unique and that they will be realized at the same time ``as everybody else's algorithms." 

\begin{figure*}[t!]
    \centering
    \includegraphics[width=\textwidth]{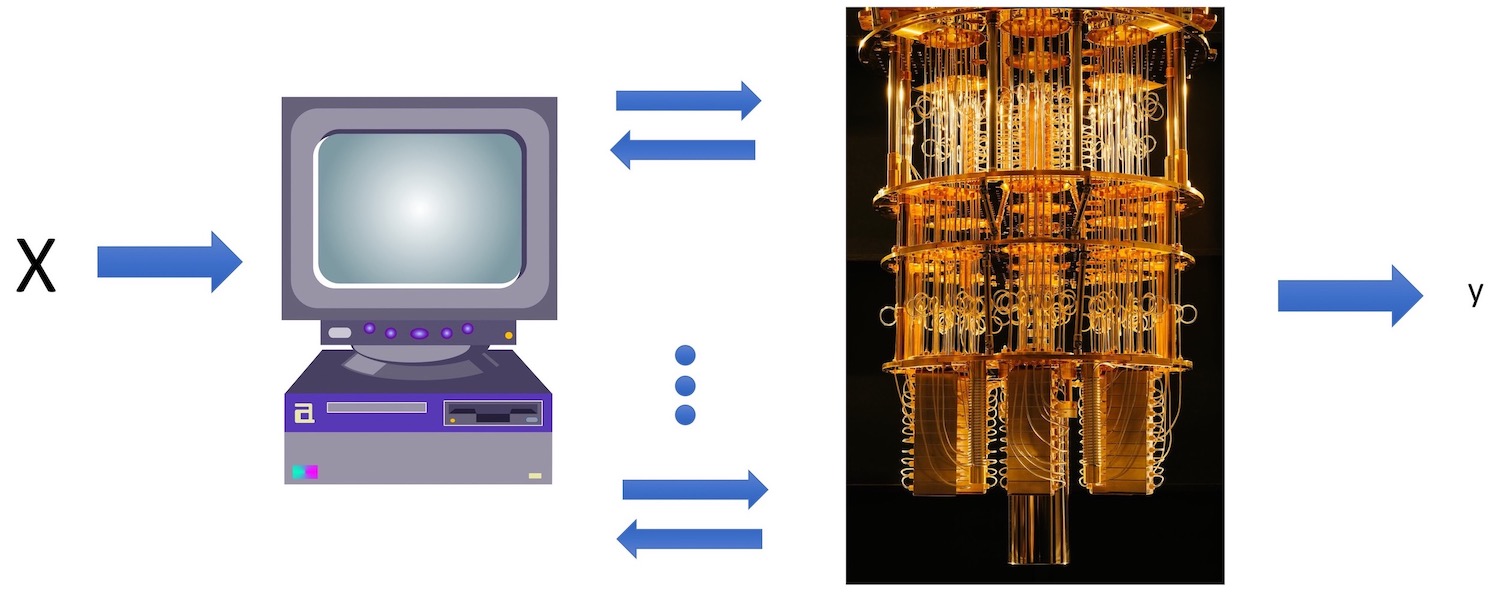}
    \caption*{A schematic of general hybrid classical-quantum algorithms, relevant
to Prof. Aram Harrow's work on hybrid optimization algorithms. Credit: Aram Harrow, IBM Corporation}
\end{figure*}

When we asked about his thoughts on particular quantum hardware modalities, Harrow noted that ``all the architectures seem to have trouble scaling." He believes that modularity will be integral to the success of any architecture. If there are many modules, he claims that making them small will also be important. Professor Harrow noted that both superconducting qubits and ion traps have candidate solutions to modularity, via photon communication, but neither technology works particularly great at the moment. ``Right now, things work much better within a module than across modules. Overall, I don't have a great insight into which one will win out. Each one is going to need some more breakthroughs. It's not totally crazy to imagine that even a current underdog like linear optics might prevail, but right now ions and superconducting qubits are the clear leaders."

\subsection*{What are the big challenges in quantum information?}

Professor Harrow believes that the field currently faces a lot of big challenges. In particular, he feels that ``theorists can be a bit wasteful in terms of the number of qubits [they] ask for." It is important that theorists find ways of doing things like error correction and fault-tolerance, while making fewer demands of experimentalists. However, Harrow believes that the biggest challenge is actually that ``classical computers are very good." This means that quantum computing researchers need to find ways to compete with classical computers for useful problems. Part of the challenge in this is that we do not actually know how to prove that certain operations will work on classical computers. In particular, Professor Harrow discussed an example of gradient descent on functions that are not convex. These functions have several local minima and, although gradient descent is not guaranteed to give a good answer, in practice it often does. ``In quantum computing, because we have not been able to test our algorithms, we try to prove things, but you can often prove a lot less than you could just have a guess for and test. So I think in algorithm design, we have to figure out more heuristics." Generally speaking, Harrow believes we do not yet have a great sense of how to use quantum computers. He notes that we solely have a few ideas of how to use them for applications, like simulating molecules. Thus, he believes that a lot of work remains in coming up with better algorithms and improving the performance of currently existing algorithms. 

Professor Harrow also believes that a lot of interesting questions remain in the domain of ``applying quantum information as a lens to the rest of physics." He notes that people have already begun to pursue such work in the domains of the black hole information problem, quantum phase transitions, and topological order in many-body systems. ``In general, there's a lot of promise for using quantum information to think about other topics in physics and I think there's a lot more to be done there." Harrow himself has pursued work along this line. He pointed out that his previously discussed work on mixing entanglement connects to rigorously proving the validity of the Mean Field Approximation in systems where one particle interacts with many other particles. ``You know, a lot of times you are doing work in basic quantum information theory and something like this will come out of it."

\subsection*{What question is Prof. Harrow currently looking to solve?}
One problem that Professor Harrow is currently interested in, is figuring out how a small quantum computer could be helpful for a practical machine learning or optimization problem. In fact, he has proposed a strategy for how a quantum computer can meaningfully interact with a database that is far too large for a quantum computer to read. The solution, as it turns out, is working together with a classical computer. Thus, Harrow proposes a \textit{\color{DarkRed}hybrid algorithm}, in which a quantum and classical computer are used together, each doing something that the other cannot do. He claims to have a few ideas and algorithms under works, but believes there is still a lot of room for progress. ``Of course, we don't understand classical machine learning fully. We just have a collection of algorithms and we don't have a general sense of how well they work. We've just seen them work on a bunch of examples and we start to gain some confidence. So, I think it's too early to hope to have one solution for all the quantum cases, but I think that would be an important area for more progress." 

Harrow claims that if we want quantum computers to be useful, we ``have to go where the hard computing problems are. Otherwise, why bother with the effort to make a quantum computer?" He notes that we understand a lot of these problems, such as code breaking, pretty well. There are other areas, such as quantum simulation, that we do not yet fully understand, but we have made a lot of progress on. However, he believes the domain in which we currently spend a lot of computing resources is optimization and machine learning.  ``I think there's just a ton of uncertainty as to how useful a quantum computer would be in this space. And so, it's not like there's one big result I would hope for that would solve it, but I think there's opportunity for a lot of progress, both in terms of theoretical progress (coming up with new algorithmic ideas and frameworks) and also really concrete things (like saying for this dataset here's an algorithm we could try and here's some idea of why we expect it to work better)." Harrow is hopeful that he could even try out approaches to these problems on a near-term quantum computer or simulate them on classical computers. This, he believes, would allow researchers to demonstrate that, given a quantum computer, they could do a much better job than solely with access to a classical device. ``So, that's something where there's room for a lot of progress and I think it would be useful to make that progress."

\subsection*{What exactly is hybrid quantum-classical computing?}
Since the early days of quantum computing research, scientists have proposed making quantum computers fault tolerant by utilizing the increased reliability of classical computers. ``So, the classical  computer would measure some qubits and then figure out what the error was and how to fix it. And that would be a very simple quick calculation for the classical computer." More recently, researchers have proposed hybrid quantum-classical computers for \textit{\color{DarkRed}variational quantum algorithms}. This type of algorithm makes use of a quantum circuit, or a sequence of quantum operations, with a number of free parameters. It is very quick to run the quantum circuit on a quantum computer several times and the resulting outputs are fed into a classical computer. The classical computer then decides how to tune the free parameters of the quantum circuit, so as to achieve a desired output. ``That's another example of a hybrid algorithm. And again, you're not making use of any computational heavy lifting by the classical computer, you're really just making use of the fact that it can store something for a long time. Keeping track of where you are in the gradient descent and not being subject to decoherence is something where a classical computer is more useful."

Professor Harrow claims that while ``a lot of people are thinking about hybrid computing in various forms, the form [he is] proposing is a little bit different." Specifically, in his view of hybrid computing, the role of a classical computer is different from the usual sense. In particular, he makes use of a different feature of a quantum computer, ``which is that it has a big hard drive and can access a lot of data." Although Harrow is not new in proposing hybrid quantum computing, he believes this aspect is one that ``has not been looked at so much before and should be looked at more."

\subsection*{What kinds of problems will we use quantum computers to solve?}

Professor Harrow is of the firm belief that cryptanalysis will be an important application of quantum computers in the short-run, but not in the long-run. According to Harrow, we all currently use codes that are vulnerable to quantum attack. However, he believes we will soon transition to new, quantum-robust codes and ``then people will not spend much time running quantum computers to break codes, because those codes won't be in widespread use." However, he notes that it will probably take a long time to turn over our current cryptosystems, leaving us in a period of vulnerability. When we asked Professor Harrow if there already existed a code that is completely quantum robust, he responded that ``nothing is ever for sure." In fact, ``we don't even know whether our current cryptosystems are vulnerable to classical attack. There could be some classical algorithm tomorrow that could break everything." He explained, however, that there is an entire field of research dedicated to so-called \textit{\color{DarkRed}post-quantum cryptography}. Theorists in the field have developed several proposals of cryptosystems that appear resilient to quantum attack. However, Harrow claims that these cryptosystems are less practical because the legitimate party has to spend a lot more resources in encoding and decoding the key. ``Our current key size might be 1000 bits, but one of these other systems might take a key size of a million bits. So that's a little more resource intensive for your phone to use.”"

An application that Professor Harrow seemed far more excited about was quantum simulation. ``I think this is more promising because, you know, benzene is never going to get more complicated. It will not change its structure just because we found a new way of analyzing it." In addition to chemistry, Harrow notes that quantum simulation could apply to nuclear physics, high energy physics, and condensed matter physics. In general, it will apply ``to many problems where our current theoretical and numerical approaches have gotten stuck." Professor Harrow believes there is still a lot of work to be done in this domain, ``because existing techniques are pretty good and there's a lot to compete with." However, in the long-run, he is confident this will be a worthwhile effort. ``You can tell because we spend a lot of money doing this right now. Pharmaceutical companies pay a lot of money to simulate biological molecules, to try and figure out what they will do, just on the computer. Academic researchers spend a lot of computing cycles simulating the proton and there are many things like this where we are spending a lot of computer time to get a very crude answer. So there's a very clear demand for improved answers and for extending these simulations to areas where current simulations don't even attempt." 

Finally, Professor Harrow believes that machine learning and optimization are fields which would have even wider use. The economic and social benefit could be even greater. However, he feels that the advantage of quantum over classical algorithms for this is much more uncertain.

\subsection*{Thoughts on current media hype?}
Professor Harrow believes that current hype surrounding quantum computing is both justified and not. He notes that while there has been great progress in the field, it has been very steady. ``The interest in the field has dramatically jumped, but the progress has not been so dramatic. It's just been steadily plodding along." While Harrow believes the progress has been very good from a scientific standpoint, he worries that general expectations may be too high. ``It's not like next year everything is going to be solved. We're probably just going to keep on incrementally moving forward. It's going to be exciting, but it will still take a while for it all to shake out." As was the case with classical computing, Harrow believes that there is going to be an increasing need for quantum engineering, ``which is really a field that's in its infancy.” Harrow does not believe the increase in hype means that we have achieved our goals. “It just means our hope in them is feeling more justified."

\bibliographystyle{unsrt}
\bibliography{bibliography}

\end{document}